## OPTICAL SYSTEMS

# A New Demultiplexer Based Upon Two-Dimensional Photonic Crystals for Optical Integrated High-Density Circuits [*]

**I.V. Guriev, I.A. Sukhoivanov, A.V. Shulika, and A.V. Kublik**

Kharkov National University of Radio Engineering and Electronics,
14, Lenin Ave, Kharkiv, 61166, Ukraine

**ABSTRACT:** A new conception of the wavelength division multiplexing on the basis of the two-dimensional photonic crystal was developed. The photonic crystals band structures were investigated in details. The method for the synthesis of the two-channel wavelength division multiplexer on the basis of two dimensional photonic crystals was developed using this method the device was synthesized. The spectral characteristics of the device and its dynamic pulse pattern response were investigated.

## INTRODUCTION

At present photonic crystals (PhC) that are characterized by a periodic variation in a refractive index offer a fertile field for research. The matter is that these crystals have opened up fresh opportunities for controlling wave propagation over the range from millimeter wave to IR and visible spectral region [1,2].

Photonic crystal is an area of great interest in terms of transmitting and processing some information in fully optical systems because a photonic forbidden band originate an radiation is localized in the region of periodic-structure defects. Specifically, from among the critical areas where the PhCs find their use, one can refer not only to low threshold lasers that act as a factor of optical constraint but also to the filtering system for transmitting a definite optical spectrum in optical waveguides and devices intended for wavelength-

---

 

division multiplexing systems [3,4], in multiplexers and combiners [5] based on the photonic crystals.

It is matter of general knowledge that by now the PhC filters can constructed by varying photonic forbidden band (FB) of a structure [3] in 1-, 2- and 3-D photonic crystals [6-8]. Taking a different approach, the filters, in general, and demultiplexers, in particular, can be built around 1-D photonic crystals, i.e., Bragg reflectors [9] in combination with circulators [10]. These types of multiplexers represent a sequentially arranged circulators alternating with Bragg reflectors being turned to definite wavelengths. This technology makes it possible to multiplex the frequency channels with a spacing of 100GHz. However, for a relatively high efficiency of channel splitting, these channels differ in rather large dimensions.

In the present paper we have looked into a novel wavelength-division multiplexing technique using the wide-band 2-D PhC filters. A wavelength-division multiplexing design based upon the integrated-type photonic crystal is proposed. The above construction is a waveguide channel made by introducing a defect into a periodic structure of a 2-D photonic crystal. The end of the channel of the same structure is fitted with a 3dB power splitter. At the input of each of power splitter ports a wide-band filter is installed, which is made through variation of geometric PhC parameters. Each of the filters has a forbidden band corresponding to a strictly calculated spectral range. Thus, the wavelength channel splitting can be performed efficiently.

In the present paper an in-depth analysis of the 2-D photonic crystals having a cubic and hexagonal crystal lattice and round-shaped elements is performed. The novel channel wavelength-division devices based upon wide-band frequency filters have been developed and their parameters calculated. The spectral characteristics of such devices have been studied comprehensively and optimized and a study of their information properties have been made.

## DESCRIPTION OF THE MODEL

From what has been said an optical channel divider is the photonic crystal in which the conditions are provided for selective radiated-wave propagation by implanting defects into a strictly periodic structure and by changing the parameters of elements. These conditions can be investigated through the analysis of forbidden bands of different photonic crystal regions.

An infinite, strictly periodic photonic crystal structure is featured by a photonic forbidden band that depends upon the relation between the refractive indexes of elements and the basic substance as well as upon the structure geometric parameters. While bilding the afore-mentioned device by modifying the structure we disturb the strict periodicity that makes up the regions having an altered photonic forbidden band. As we proceeded with some reasoning the preliminary calculations led us to assume that the photonic forbidden band of the infinite structure does not differ qualitatively from the forbidden band of the finite-dimensions structure. This consideration is found to be valid, as it is corroborated by the following numerical experiment on studying the structure characteristics. Thus, in order to calculate a forbidden band from each of the device's regions one can resort to the methods applicable to an infinite, strictly periodic structure.

To obtain PhC zone structure it is necessary to solve the eigenvalue problem for the Helmholtz wave equations (1). For the given problem we make use of the Helmholtz steady-state equations in an approximation of nonmagnetic media:

$$\frac{1}{\varepsilon(\vec{r})}\nabla\times\left\{\nabla\times\vec{E}(\vec{r})\right\}+\frac{\omega^2}{c^2}\vec{E}(\vec{r})=0,$$

(1)

$$\nabla\times\left\{\frac{1}{\varepsilon(\vec{r})}\nabla\times\vec{H}(\vec{r})\right\}+\frac{\omega^2}{c^2}\vec{H}(\vec{r})=0,$$

where $\vec{E}$ and $\vec{H}$ are the electric and magnetic field intensities respectively, $\vec{r}$ is the generalized coordinate, $\omega$ is the eigenfrequency of the structure; $c$ is the speed of light in vacuum; $\varepsilon(\vec{r})$ is the profile of relative permittivity of a medium.

In order to find the eigenvalues of the structure's frequency it would be convenient to make use of the Bloch theorem to represent functions $\vec{E}(\vec{r})$ and $\vec{H}(\vec{r})$ as a plane wave multiplied by the periodic function with a structure period:

$$\vec{E}(\vec{r})=\vec{u}_{\vec{k},n}(\vec{r})e^{i\cdot\vec{k}\cdot\vec{r}},$$

(2)

$$\vec{H}(\vec{r})=\vec{v}_{\vec{k},n}(\vec{r})e^{i\cdot\vec{k}\cdot\vec{r}},$$

where $\vec{u}_{\vec{k},n}(\vec{r})$ and $\vec{v}_{\vec{k},n}(\vec{r})$ are the periodic functions, $\vec{k}$ is the wave vector, $n$ is the number of a band.

Function $\frac{1}{\varepsilon(\vec{r})}$, then $\vec{E}(\vec{r})$ and $\vec{H}(\vec{r})$ are expanded into a Fourier series in terms of the reciprocal lattice:

$$\frac{1}{\varepsilon\left(\vec{r}\right)} = \sum_{\vec{G}} \chi\left(\vec{G}\right)\exp\left(i\vec{G}\cdot\vec{r}\right),$$

$$\vec{E}_{\vec{k},n}\left(\vec{r}\right) = \sum_{\vec{G}}\vec{E}_{\vec{k},n}\left(\vec{G}\right)\exp\left(i\cdot\left(\vec{k}+\vec{G}\right)\cdot\vec{r}\right), \qquad (3)$$

$$\vec{H}_{\vec{k},n}\left(\vec{r}\right) = \sum_{\vec{G}}\vec{H}_{\vec{k},n}\left(\vec{G}\right)\exp\left(i\cdot\left(\vec{k}+\vec{G}\right)\cdot\vec{r}\right),$$

where $\vec{G}$ is the reciprocal lattice vector. Substituting (3) into (1) we arrive at a set of linear equations for $\omega_{\vec{k},n}$, i.e., the eigenfrequencies at a given point of $k$-space:

$$-\sum_{\vec{G}}\chi\left(\vec{G}-\vec{G}'\right)\left(\vec{k}+\vec{G}\right)\times\left\{\left(\vec{k}+\vec{G}'\right)\times\vec{E}_{\vec{k},n}\left(\vec{G}'\right)\right\} = \frac{\omega_{\vec{k},n}}{c^2}\vec{E}_{\vec{k},n}\left(\vec{G}'\right),$$

$$\sum_{\vec{G}}\chi\left(\vec{G}-\vec{G}'\right)\left(\vec{k}+\vec{G}\right)\times\left\{\left(\vec{k}+\vec{G}'\right)\times\vec{H}_{\vec{k},n}\left(\vec{G}'\right)\right\} = \frac{\omega_{\vec{k},n}}{c^2}\vec{H}_{\vec{k},n}\left(\vec{G}'\right).$$

(4)

As we solve one of the given systems (1.4) for each point of $k$-space in the first Brillouin zone by a numerical procedure, we get the photonic crystal zone structure.

The key shortcomings of this particular method are as follows:

1. It is intact impossible to allow for the losses in a substance. This shortcoming results from the assumption that in wave equations the operators are found to be linear only if the refractive index of a medium does not contain a complex portion.

2. It is impossible to take account of the dispersion of a material. To find the frequency eigenvalues, it is necessary that the refractive index distribution be unambiguously preasigned. It is evident that in this instance there is no way of taking into account the material dispersion.

3. The calculation time is much too extended with an accuracy being low, in which case this, however, is compensated by as relative ease of implementation.

There are two ways of constructing the devices whose principle of operation is based upon the variation of the photonic crystal forbidden bands: the first is to bring about the change in the relations of refractive indexes, with the structure

484

configuration remaining intact, the second is to change the structure geometry at constant refractive indexes.

Figure 1(a) shows the dependence of the position of the forbidden band and its width upon the radii of rods. The relative frequency laid off on the ordinate axis is determined from the formula:

$$\omega_r = \frac{\omega \cdot a}{2 \cdot \pi \cdot c} = \frac{a}{\lambda}. \tag{5}$$

Here we have to deal with the following constant parameters: the refractive index of the basic substance $n_1 = 1$, the refractive index of rods $n_2 = 3$. The spacing between the elements $a = 0.55\,\mu$ m. The ratio of the element's radii to the interelement spacing varies from 0.1 to 0.5.

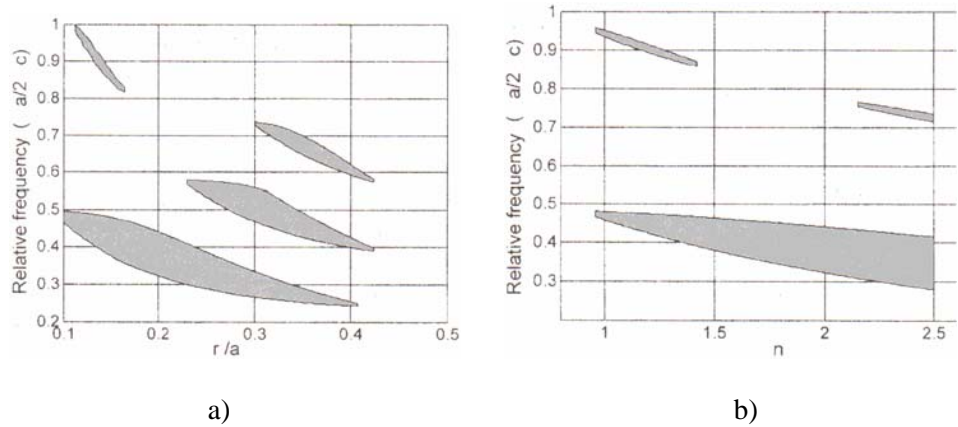

a)                                                    b)

**FIGURE 1.**

As will be seen from the Fig. 1, the lower forbidden band is the most extended one. As the radius of the elements increases, the corresponding frequencies tend to decrease. At the same time its width changes likewise. In this case the widest forbidden band is observed in a region where the relation between the element's radius and the interelement spacing is equal to 0.2.

Figure 1(b) presents the dependence of the position and width of the forbidden band upon the relation between the refractive index of the basic substance and the components. The value of $\Delta n$, i.e., the difference between the refractive index of elements and the basic substance is plotted on the abscissa axis. The constant parameters of the structure are therewith as follows: the refractive index of the basic substance $n_1 = 1$, the distance between the centers of rods $a = 0.55\,\mu$ m, the ratio of the elements' radius to the interelement spacing is 0.1. As evident from Fig.1, the width of the forbidden band shows a monotonous growth with increasing refractive index differences. As it takes place, the forbidden band center is shifted towards the low frequency range.

Figure 1 demonstrates the examples of how the characteristics of crystal structures are measured by changing the ratio of rods' radius to spacing between them. Well as by varying the relationship between the refractive indexes of a medium and elements. In terms of adaptability to streamlined manufacture, it would be inconvenient to make changes in the refractive index relation. It seems far simpler to alter the elements' geometric dimensions. At the same time, as seen from Fig. 1, as the ratio of elements' radius to the interelement spacing varies, there comes about the regions on the abscissa axis, where the rate at which the frequency of the upper edge of the forbidden band tends to vary is equal to that of variation in the frequency of the lower edge (see Fig. 1(a)). Given the changing differences in refractive indexes, the upper edge of the forbidden band changes very slowly. It is exactly for this reason that the above mechanism makes it difficult to construct a device in terms of bringing about the changes in the elements' refractive indexes.

From the above discussion of follows that a radius of the structure elements has been chosen as a variable parameter. In order to specify the device's key parameter it is required that the refractive indexes relation and the geometric parameters of the photonic crystals of the waveguide channels and filters be precisely present. For this purpose consider the dependence of the parameters of the PhC's forbidden band upon the element's radius – interelement spacing relation for the fixed value of refractive index relations (Fig. 2). To construct a demultiplexer a lower forbidden band has been opted (Fig. 1), because it overlaps the widest frequency range.

In Fig. 2 the normalized channel frequencies are designated by horizontal dark lines, whereas the radii of rods of the first filter, the waveguide channel and the second filter are labeled respectively by vertical lines. The PhC with a radius of elements $r_2$ is seen to have a forbidden band comprising both channels, whereas the PhC with radii of elements $r_1$ and $r_3$ have a forbidden band for one of the channels only. Since a normalized frequency is plotted on the vertical axis, the parameters can be matched for any relation of channels' wavelengths. In this paper a study has been made of the demultiplexer of 1.55 and 1.31 $\mu$m – wavelength channels based upon the cubic-lattice photonic crystal.

The procedure for selecting the structure parameters is as follows. Initially one of the relative frequencies has to be specified. Knowing a particular wavelength and a relative frequency it is possible to determine parameter $a$, i.e., the interelement spacing. Given the value of this parameter and the second wavelength one can determine the value of the second relative frequency. The frequencies should be chosen so that at a certain value of the elements' radius both of the frequencies are within the forbidden band, and this is just what corresponds to the PhC of the waveguide channel, whereas for two other values of the radius only one of the frequencies could get into the forbidden band, and this corresponds to the PhC of frequency filters. The problem can get the considerably simplified solution by a graphical method.

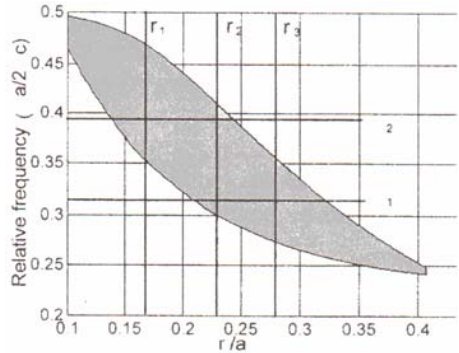

**FIGURE 2.**

In the long run we obtain the structure for the cubic-lattice photonic crystal having the following parameters: refractive index of rods $n_r = 3.5$; a distance between the rods' centers $a = 0.55\mu$ m; $r_1 = 0.108 \cdot a$; $r_2 = 0.175 \cdot a$; and $r_3 = 0.229 \cdot a$.

Thus, we can avail ourselves of the data needed to carry out the investigations into the device's spectral responses (signal power dependencies at each of the outputs) and the information characteristics (investigation of the device's maximum speed resulting from the dispersion of selective elements).

## INVESTIGATION RESULTS

A proposed two-channel demultiplexer is a device developed around a 2-D photonic cubic- and hexagonal-lattice photonic crystal. This demultiplexer is intended for demultiplexing the 1.31 and 1.55 $\mu$ m wavelength channels. Wavelength demultiplexing is performed by using the splitters [5,10,11]. We propose that wide-band filters formed by 2-D photonic crystals (their geometric parameters differ from those of basic parameters) be installed in output splitter channels.

The characteristics of such a dimultiplexer have been examined through a numerical experiment involving the finite-difference method [12-14]. The results obtained from calculating the electromagnetic field distribution are presented in Fig. 3.

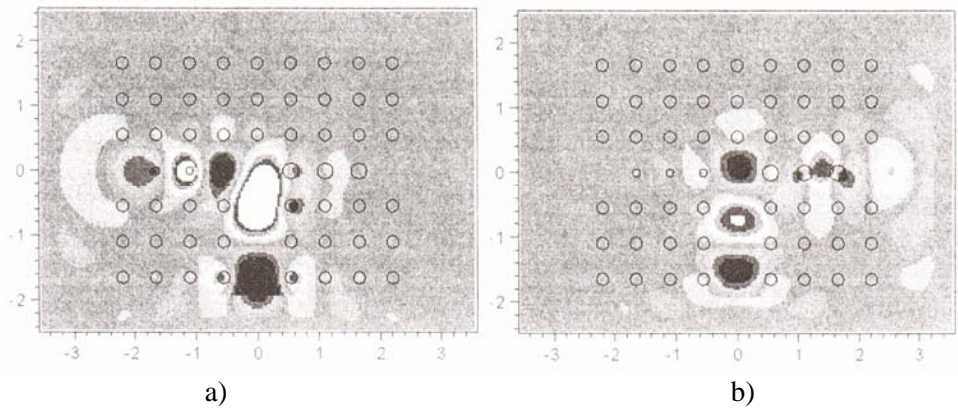

a)                               b)

**FIGURE 3.**

The efficiency of splitting the frequency channels can be analyzed by plotting spectral characteristic curves. Figure 4 depicts these characteristics of the devices based upon the photonic cubic-lattice crystal. The calculation of these curves reduced to deriving transfer characteristics for each of the wavelengths.

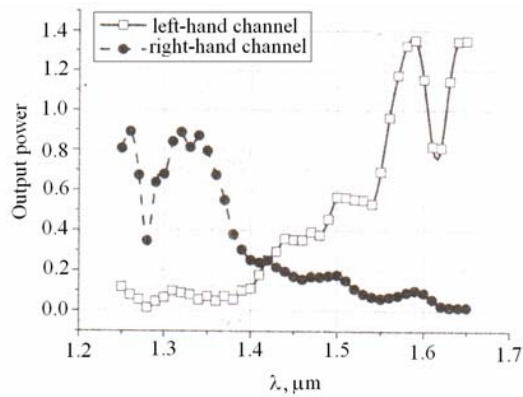

**FIGURE 4.**

As will be apparent from Fig. 4, the spectral characteristic have identical transmission values at operating wavelengths, thereby ensuring that the power between the demultiplexer channels is uniform.

## CONCLUSIONS

The present paper has discussed the properties of photonic forbidden-band structures. A technique of expanding into plane waves is scrutinized to obtain the band-like diagrams of structures having a periodic variation in a parameter, particularly, of a photonic crystal. A novel wavelength demultiplexer construction is proposed in term of a 2-D integration-type photonic crystal. The principles of operation of the afore-mentioned devices have studied extensively and substantiated. Besides, their basic properties have been investigated at length and appropriate inferences made.

The results thus presented indicated that the proposed concept of constructing integral optical demultiplexers can be adequately utilized in short-pulse optical systems as well as in existing high-speed data transmission systems.